# Discrete, 3D distributed linear imaging methods of electric neuronal activity. Part 1: exact, zero error localization

Roberto D. Pascual-Marqui
The KEY Institute for Brain-Mind Research
University Hospital of Psychiatry
Lenggstr. 31, CH-8032 Zurich, Switzerland
pascualm at key.uzh.ch
www.keyinst.uzh.ch/loreta

## 1. Abstract

This paper deals with the EEG/MEG neuroimaging problem: given measurements of scalp electric potential differences (EEG: electroencephalogram) and extracranial magnetic fields (MEG: magnetoencephalogram), find the 3D distribution of the generating electric neuronal activity. This problem has no unique solution. Only particular solutions with "good" localization properties are of interest, since neuroimaging is concerned with the localization of brain function. In this paper, a general family of linear imaging methods with exact, zero error localization to point-test sources is presented. One particular member of this family is sLORETA (standardized low resolution brain electromagnetic tomography; Pascual-Marqui, Methods Find. Exp. Clin. Pharmacol. 2002, 24D:5-12; http://www.unizh.ch/keyinst/NewLORETA/sLORETA/sLORETA-Math01.pdf). It is shown here that sLORETA has no localization bias in the presence of measurement and biological noise. Another member of this family, denoted as eLORETA (exact low resolution brain electromagnetic tomography; Pascual-Marqui 2005: http://www.research-projects.unizh.ch/p6990.htm), is a genuine inverse solution (not merely a linear imaging method) with exact, zero error localization in the presence of measurement and structured biological noise. The general family of imaging methods is further extended to include data-dependent (adaptive) quasi-linear imaging methods, also with the exact, zero error localization property.

## 2. The forward equation

Details on the electrophysiology and physics of EEG/MEG generation can be found in Mitzdorf (1985), Llinas (1988), Martin (1991), Hämäläinen et al (1993), Haalman and Vaadia (1997), Sukov and Barth (1998), Dale et al (2000), Baillet et al. (2001). The basic underlying physics can be studied in Sarvas (1987).

Consider the forward EEG equation:

**Eq. 1:** $\quad \mathbf{\Phi} = \mathbf{KJ} + c\mathbf{1}$

where the vector $\mathbf{\Phi} \in \mathbb{R}^{N_E \times 1}$ contains instantaneous scalp electric potential differences measured at $N_E$ electrodes with respect to a single common reference electrode (e.g., the reference can be linked earlobes, the toe, or one of the electrodes included in $\mathbf{\Phi}$); the matrix $\mathbf{K} \in \mathbb{R}^{N_E \times (3N_V)}$ is the lead field matrix corresponding to $N_V$ voxels; $\mathbf{J} \in \mathbb{R}^{(3N_V) \times 1}$ is the current density; $c$ is a scalar accounting for the physics nature of electric potentials which are determined up to an arbitrary constant; and **1** denotes a vector of ones, in this case $\mathbf{1} \in \mathbb{R}^{N_E \times 1}$. Typically $N_E \ll N_V$, and $N_E \geq 19$.

In Eq. 1, the structure of **K** is:

**Eq. 2:**
$$\mathbf{K} = \begin{pmatrix} \mathbf{k}_{11}^T & \mathbf{k}_{12}^T & \ldots & \mathbf{k}_{1N_V}^T \\ \mathbf{k}_{21}^T & \mathbf{k}_{22}^T & \ldots & \mathbf{k}_{2N_V}^T \\ \ldots & & & \\ \mathbf{k}_{N_E 1}^T & \mathbf{k}_{N_E 2}^T & \ldots & \mathbf{k}_{N_E N_V}^T \end{pmatrix}$$

where the superscript "$T$" denotes transposition; and $\mathbf{k}_{ij} \in \mathbb{R}^{3\times 1}$, for $i=1\ldots N_E$ and for $j=1\ldots N_V$, corresponds to the scalp potentials at the $i$-th electrode due to three orthogonal unit strength dipoles at voxel $j$, each one oriented along the coordinate axes $x$, $y$, and $z$. For instance, in infinite homogeneous medium with conductivity $\sigma$:

**Eq. 3:**
$$\mathbf{k}_{ij} = \frac{1}{4\pi\sigma} \frac{(\mathbf{r}_{Ei} - \mathbf{r}_{Vj})}{\|\mathbf{r}_{Ei} - \mathbf{r}_{Vj}\|^3}$$

where $\mathbf{r}_{Ei}, \mathbf{r}_{Vj} \in \mathbb{R}^{3\times 1}$ are position vectors for the $i$-th scalp electrode and for the $j$-th voxel, respectively. As another example, for the case of a homogeneous conducting sphere in air, the lead field is:

**Eq. 4:**
$$\mathbf{k}_{ij} = \frac{1}{4\pi\sigma} \left[ 2\frac{(\mathbf{r}_{Ei} - \mathbf{r}_{Vj})}{\|\mathbf{r}_{Ei} - \mathbf{r}_{Vj}\|^3} + \frac{\mathbf{r}_{Ei}\|\mathbf{r}_{Ei} - \mathbf{r}_{Vj}\| + (\mathbf{r}_{Ei} - \mathbf{r}_{Vj})\|\mathbf{r}_{Ei}\|}{\|\mathbf{r}_{Ei}\|\|\mathbf{r}_{Ei} - \mathbf{r}_{Vj}\|\left[\|\mathbf{r}_{Ei}\|\|\mathbf{r}_{Ei} - \mathbf{r}_{Vj}\| + \mathbf{r}_{Ei}^T(\mathbf{r}_{Ei} - \mathbf{r}_{Vj})\right]} \right]$$

In the previous equations, the following notation was used:

**Eq. 5:** $\quad \|\mathbf{X}\|^2 = tr(\mathbf{X}^T \mathbf{X}) = tr(\mathbf{X}\mathbf{X}^T)$

where $tr$ denotes the trace, and **X** is any matrix or vector. If **X** is a vector, then this is the squared Euclidean $L_2$ norm; if **X** is a matrix, then this is the squared Frobenius norm.

Note that **K** can also be conveniently written as:

**Eq. 6:** $\quad \mathbf{K} = (\mathbf{K}_1, \mathbf{K}_2, \mathbf{K}_3, \ldots, \mathbf{K}_{N_V})$

where $\mathbf{K}_j \in \mathbb{R}^{N_E \times 3}$, for $j=1\ldots N_V$, is defined as:

**Eq. 7:**
$$\mathbf{K}_j = \begin{pmatrix} \mathbf{k}_{1j}^T \\ \mathbf{k}_{2j}^T \\ \ldots \\ \ldots \\ \mathbf{k}_{N_E j}^T \end{pmatrix}$$

Ideally, the lead field should correspond to the real head (with realistic geometry and conductivities). For the EEG problem, the voxels should correspond to cortical grey matter. For other situations (e.g. EKG), appropriate volume conductor models and solution spaces should be used.

In Eq. 1, **J** is structured as:

Eq. 8:
$$\mathbf{J} = \begin{pmatrix} \mathbf{j}_1 \\ \mathbf{j}_2 \\ \dots \\ \dots \\ \mathbf{j}_{N_V} \end{pmatrix}$$

where $\mathbf{j}_i \in \mathbb{R}^{3 \times 1}$ denotes the current density at the *i*-th voxel.

## 3. The reference electrode problem

As a first step, before even stating the inverse problem, the reference electrode problem will be solved, by estimating "*c*" in Eq. 1. Given $\mathbf{\Phi}$ and $\mathbf{KJ}$, the reference electrode problem is:

Eq. 9: $$\min_c \|\mathbf{\Phi} - \mathbf{KJ} - c\mathbf{1}\|^2$$

The solution is:

Eq. 10: $$c = \frac{\mathbf{1}^T}{\mathbf{1}^T \mathbf{1}} (\mathbf{\Phi} - \mathbf{KJ})$$

Plugging Eq. 10 into Eq. 1 gives:

Eq. 11: $$\mathbf{H\Phi} = \mathbf{HKJ}$$

where:

Eq. 12: $$\mathbf{H} = \mathbf{I} - \frac{\mathbf{1}\mathbf{1}^T}{\mathbf{1}^T \mathbf{1}}$$

is the average reference operator, also known as the centering matrix, and $\mathbf{I} \in \mathbb{R}^{N_E \times N_E}$ is the identity matrix.

This result establishes the fact that any inverse solution (of any form, not necessarily linear) will not depend on the reference electrode.

Henceforth, it will be assumed that the EEG measurements and the lead field are average reference transformed, i.e.:

Eq. 13: $$\begin{cases} \mathbf{\Phi} \leftarrow \mathbf{H\Phi} \\ \mathbf{K} \leftarrow \mathbf{HK} \end{cases}$$

and Eq. 1 will be rewritten as:

Eq. 14: $$\mathbf{\Phi} = \mathbf{KJ}$$

Note that **H** plays the role of the identity matrix for EEG data. It actually is the identity matrix, except for a null eigenvalue corresponding to an eigenvector of ones (see Eq. 12), accounting for the reference electrode constant.

## 4. A family of discrete, 3D distributed linear imaging methods with exact, zero error localization

The family of linear imaging methods considered here is parameterized by a symmetric matrix $\mathbf{C} \in \mathbb{R}^{N_E \times N_E}$, such that:

**Eq. 15:** $$\hat{\mathbf{j}}_i = \left[ \left( \mathbf{K}_i^T \mathbf{C} \mathbf{K}_i \right)^{-1/2} \mathbf{K}_i^T \mathbf{C} \right] \boldsymbol{\Phi}$$

where $\hat{\mathbf{j}}_i \in \mathbb{R}^{3 \times 1}$ is any estimator for the electric neuronal activity at the *i*-th voxel, not necessarily current density (e.g. it can be standardized current density, as in Pascual-Marqui 2002).

Note that in the case of MEG, $\mathbf{C}$ must be non-singular. In the case of EEG, $\mathbf{C}$ must be of rank $(N_E - 1)$, with its null eigenvector equal to a vector of ones (accounting for the reference constant).

Note that in Eq. 15, the symmetric matrix $\left( \mathbf{K}_i^T \mathbf{C} \mathbf{K}_i \right)$ is of dimension $3 \times 3$, and the notation $\left[ \left( \mathbf{K}_i^T \mathbf{C} \mathbf{K}_i \right)^{-1/2} \right]$ indicates the symmetric square root inverse. In the particular case of MEG in a spherical head model, the matrix $\left( \mathbf{K}_i^T \mathbf{C} \mathbf{K}_i \right)$ is of rank two, and its symmetric square root pseudo-inverse must be used.

Localization inference in neuroimaging is typically based on the search for large values of the power (squared amplitude) of the estimator for electric neuronal activity, i.e. $\left\| \hat{\mathbf{j}}_i \right\|^2$.

In order to test the localization properties of a linear imaging method, consider the case when the actual source is an arbitrary point-test source at the *j*-th voxel. This means that:

**Eq. 16:** $$\boldsymbol{\Phi} = \mathbf{K}_j \mathbf{A}$$

where $\mathbf{K}_j$ is defined in Eq. 7 above, and $\mathbf{A} \in \mathbb{R}^{3 \times 1}$ is an arbitrary non-zero vector (containing the dipole moments).

Plugging Eq. 16 into Eq. 15 and taking the squared amplitude gives:

**Eq. 17:** $$\left\| \hat{\mathbf{j}}_i \right\|^2 = \mathbf{A}^T \mathbf{K}_j^T \mathbf{C} \mathbf{K}_i \left( \mathbf{K}_i^T \mathbf{C} \mathbf{K}_i \right)^+ \mathbf{K}_i^T \mathbf{C} \mathbf{K}_j \mathbf{A}$$

where the superscript "+" denotes the Moore-Penrose pseudoinverse (which is equal to the common inverse if the matrix is non-singular).

Following the same type of derivations as in Greenblatt et al (2005), the derivative of $\|\hat{\mathbf{j}}_i\|^2$ in Eq. 17 with respect to $\mathbf{K}_i$ is:

Eq. 18:
$$\left\{ \begin{aligned} \frac{\partial \|\hat{\mathbf{j}}_i\|^2}{\partial \mathbf{K}_i} &= 2\mathbf{C}\mathbf{K}_j \mathbf{A}\mathbf{A}^T \mathbf{K}_j^T \mathbf{C}\mathbf{K}_i \left(\mathbf{K}_i^T \mathbf{C}\mathbf{K}_i\right)^+ \\ &\quad -2\mathbf{C}\mathbf{K}_i \left(\mathbf{K}_i^T \mathbf{C}\mathbf{K}_i\right)^+ \mathbf{K}_i^T \mathbf{C}\mathbf{K}_j \mathbf{A}\mathbf{A}^T \mathbf{K}_j^T \mathbf{C}\mathbf{K}_i \left(\mathbf{K}_i^T \mathbf{C}\mathbf{K}_i\right)^+ \end{aligned} \right\}$$

It can be easily shown that this derivative is zero when $\mathbf{K}_i$ is equal to $\mathbf{K}_j$, demonstrating that this family of methods produces exactly localized maxima to point-test sources anywhere in the brain, i.e. this family of linear imaging methods attains exact, zero error localization.

Note that the choice:

Eq. 19: $\quad \mathbf{C} = \left(\mathbf{K}\mathbf{K}^T + \alpha \mathbf{H}\right)^+$

gives the sLORETA method (Pascual-Marqui 2002), where $\alpha \geq 0$ is the regularization parameter.

Note that these results can be applied in a straightforward manner to the case where the current density orientation is known (i.e. known cortical geometry), but with unknown current density amplitude.

## 5. Unbiased localization for sLORETA

As in the previous section, consider the case when the actual source is any arbitrary point-test source at the *j*-th voxel, but now the measurements are contaminated with measurement and biological noise. This means that:

Eq. 20: $\quad \mathbf{\Phi} = \mathbf{K}_j \mathbf{A} + \varepsilon_\Phi + \mathbf{K}\varepsilon_J$

where $\varepsilon_\Phi$ represents the measurement noise and $\varepsilon_J$ the biological noise. It will be assumed that both noise sources are zero mean and independent, with covariance matrices:

Eq. 21: $\quad \text{cov}(\varepsilon_\Phi) = \sigma_\Phi \mathbf{H}$

Eq. 22: $\quad \text{cov}(\varepsilon_J) = \sigma_J \mathbf{I}$

This gives the following expected covariance matrix for the measurements:

Eq. 23: $\quad \text{cov}(\mathbf{\Phi}) = \mathbf{\Sigma}_\Phi = \mathbf{K}_j \mathbf{A}\mathbf{A}^T \mathbf{K}_j^T + \sigma_\Phi \mathbf{H} + \sigma_J \mathbf{K}\mathbf{K}^T$

The corresponding expected square amplitude then is:

Eq. 24:
$$\left\{ \begin{aligned} E\left(\|\hat{\mathbf{j}}_i\|^2\right) &= tr\left[ \left(\mathbf{K}_i^T \mathbf{C}\mathbf{K}_i\right)^{-1/2} \mathbf{K}_i^T \mathbf{C} \left(\mathbf{K}_j \mathbf{A}\mathbf{A}^T \mathbf{K}_j^T + \sigma_\Phi \mathbf{H} + \sigma_J \mathbf{K}\mathbf{K}^T\right) \mathbf{C}\mathbf{K}_i \left(\mathbf{K}_i^T \mathbf{C}\mathbf{K}_i\right)^{-1/2} \right] \\ &= tr\left[ \left(\mathbf{K}_j \mathbf{A}\mathbf{A}^T \mathbf{K}_j^T + \sigma_\Phi \mathbf{H} + \sigma_J \mathbf{K}\mathbf{K}^T\right) \mathbf{C}\mathbf{K}_i \left(\mathbf{K}_i^T \mathbf{C}\mathbf{K}_i\right)^+ \mathbf{K}_i^T \mathbf{C} \right] \end{aligned} \right\}$$

The derivative of $E\left(\left\|\hat{\mathbf{j}}_i\right\|^2\right)$ in Eq. 24 with respect to $\mathbf{K}_i$ is:

Eq. 25:
$$\left\{\begin{array}{l}\dfrac{\partial E\left(\left\|\hat{\mathbf{j}}_i\right\|^2\right)}{\partial \mathbf{K}_i} = 2\mathbf{C}\left(\mathbf{K}_j\mathbf{A}\mathbf{A}^T\mathbf{K}_j^T + \sigma_\Phi \mathbf{H} + \sigma_J \mathbf{K}\mathbf{K}^T\right)\mathbf{C}\mathbf{K}_i\left(\mathbf{K}_i^T\mathbf{C}\mathbf{K}_i\right)^+ \\ \quad -2\mathbf{C}\mathbf{K}_i\left(\mathbf{K}_i^T\mathbf{C}\mathbf{K}_i\right)^+ \mathbf{K}_i^T \mathbf{C}\left(\mathbf{K}_j\mathbf{A}\mathbf{A}^T\mathbf{K}_j^T + \sigma_\Phi \mathbf{H} + \sigma_J \mathbf{K}\mathbf{K}^T\right)\mathbf{C}\mathbf{K}_i\left(\mathbf{K}_i^T\mathbf{C}\mathbf{K}_i\right)^+\end{array}\right\}$$

It can be easily shown that the derivative in Eq. 25 is zero for the sLORETA case, when the parameter matrix is:

Eq. 26:
$$\mathbf{C} = \left(\mathbf{K}\mathbf{K}^T + \dfrac{\sigma_\Phi}{\sigma_J}\mathbf{H}\right)^+$$

and when $\mathbf{K}_i$ is equal to $\mathbf{K}_j$, thus demonstrating that sLORETA produces exactly localized maxima to point-test sources anywhere in the brain, even in the presence of noise, i.e. sLORETA is unbiased.

This new result is to be contrasted with those published by Sekihara et al (2005) and Greenblatt et al (2005). They showed that under pure measurement noise, sLORETA is biased, and only attains exact localization under ideal conditions of no noise. They did not consider the more realistic case where the brain in general is always active, as modeled here by the biological noise. Under these arguably much more realistic conditions, sLORETA is unbiased.

## 6. eLORETA: exact low resolution brain electromagnetic tomography

The eLORETA method was developed and officially recorded as a working project at the University of Zurich in March 2005. A description (including the official registration date) can be obtained from the University of Zurich server at:
http://www.research-projects.unizh.ch/p6990.htm

An additional reference to eLORETA is:
Roberto D. Pascual-Marqui, Alberto D. Pascual-Montano, Dietrich Lehmann, Kieko Kochi, Michaela Esslen, Lutz Jancke, Peter Anderer, Bernd Saletu, Hideaki Tanaka, Koichi Hirata, E. Roy John, Leslie Prichep. Exact low resolution brain electromagnetic tomography (eLORETA). Neuroimage 2006, Vol 31, Suppl. 1, page:S86

Consider the general weighted minimum norm solution (see, e.g. Pascual-Marqui 1999):

Eq. 27: $\hat{\mathbf{J}} = \mathbf{T}\boldsymbol{\Phi}$

with:

Eq. 28: $\mathbf{T} = \mathbf{W}^{-1}\mathbf{K}^T\left(\mathbf{K}\mathbf{W}^{-1}\mathbf{K}^T + \alpha \mathbf{H}\right)^+$

where $\mathbf{W} \in \mathbb{R}^{(3N_V)\times(3N_V)}$ denotes the symmetric weight matrix, and $\alpha \geq 0$ denotes the regularization parameter.

The particular case of interest here will only consider a structured block-diagonal weight matrix $\mathbf{W}$, where all matrix elements are zero except for the diagonal sub-blocks denoted as $\mathbf{W}_i \in \mathbb{R}^{3 \times 3}$, the *i*-th voxel, with $i = 1...N_V$.

Note that for $\alpha = 0$, this is a genuine solution, in the sense that $\hat{\mathbf{J}}$ is a direct estimator for the current density, and it reproduces exactly the measurements. In other words, for $\alpha = 0$:

Eq. 29:
$$\begin{cases} \mathbf{\Phi} = \mathbf{K}\hat{\mathbf{J}} = \mathbf{KT\Phi} \\ \mathbf{H} = \mathbf{KT} \end{cases}$$

The current density estimator at the *i*-th voxel then is:

Eq. 30:
$$\hat{\mathbf{j}}_i = \mathbf{W}_i^{-1} \mathbf{K}_i^T \left( \mathbf{KW}^{-1}\mathbf{K}^T + \alpha \mathbf{H} \right)^+ \mathbf{\Phi}$$

Based on the results of the previous section (entitled "*A family of discrete, 3D distributed linear imaging methods with exact, zero error localization*"), by comparing Eq. 30 with Eq. 15, exact, zero error localization is attained with weights satisfying:

Eq. 31:
$$\mathbf{W}_i = \left[ \mathbf{K}_i^T \left( \mathbf{KW}^{-1}\mathbf{K}^T + \alpha \mathbf{H} \right)^+ \mathbf{K}_i \right]^{1/2}$$

This result is easily derived by noting that Eq. 30 matches Eq. 15 when:

Eq. 32:
$$\mathbf{C} \leftarrow \left( \mathbf{KW}^{-1}\mathbf{K}^T + \alpha \mathbf{H} \right)^+$$

and:

Eq. 33:
$$\mathbf{W}_i^{-1} \leftarrow \left( \mathbf{K}_i^T \mathbf{C} \mathbf{K}_i \right)^{-1/2}$$

The weights satisfying the system of equations given by Eq. 31 define the eLORETA method, which is a genuine solution to the inverse problem (not merely a linear imaging method), and attains exact, zero error localization. Additionally, eLORETA is standardized by definition, meaning that its theoretical expected variance is unity.

Furthermore, following the derivations as in the previous section entitled "*Unbiased localization for sLORETA*", it can easily be shown that eLORETA is unbiased in the presence of measurement and structured biological noise of the form:

Eq. 34:
$$\text{cov}(\varepsilon_J) = \sigma_J \mathbf{W}^{-1}$$

Unfortunately, such a structure on background brain activity (the so-called biological noise) is determined by the physics properties of the head model and the laws of electrodynamics, and might have little relation to electrophysiological reality. This might be seen as a disadvantage of eLORETA as compared to sLORETA.

# 7. An alternative theoretical approach to eLORETA, including numerical methods

## 7.1. The classical weighted minimum norm tomography

Consider the regularized, weighted minimum norm problem:

**Eq. 35:** $$\min_{\mathbf{J}} \left[ \|\mathbf{\Phi} - \mathbf{KJ}\|^2 + \alpha \mathbf{J}^T \mathbf{WJ} \right]$$

where $\mathbf{W} \in \mathbb{R}^{(3N_V) \times (3N_V)}$ denotes a given symmetric weight matrix, and $\alpha \geq 0$ denotes the regularization parameter.

The solution is linear:

**Eq. 36:** $$\hat{\mathbf{J}} = \mathbf{T\Phi}$$

with:

**Eq. 37:** $$\mathbf{T} = \mathbf{W}^{-1} \mathbf{K}^T \left( \mathbf{K} \mathbf{W}^{-1} \mathbf{K}^T + \alpha \mathbf{H} \right)^+$$

where the superscript "+" denotes the Moore-Penrose pseudoinverse (which is equal to the common inverse if the matrix is non-singular).

The choice $\mathbf{W} = \mathbf{I}$ gives the classical minimum norm solution. This was the first 2D distributed linear solution introduced in MEG by Hämäläinen and Ilmoniemi (1984). Some of the images in that publication show that when the solution space is parallel to the measurement space, point-test sources are correctly localized, albeit with low resolution.

However, when the solution space is extended to 3D, the minimum norm solution is utterly incapable of correct localization of depth. This was clarified in Pascual-Marqui (1999), where it was shown that the minimum norm solution is harmonic, and harmonic functions attain their extreme values on the boundary of their domain of definition. This means that deep sources are always incorrectly localized to the outermost cortex.

Another popular choice is depth weighting for the 3D solution space, i.e. larger weights are assigned to deeper sources, with the hope of correcting depth localization error. These solutions achieve lower localization error than the classical minimum norm, but their errors are still significant, no matter what inverse power for depth weighting is used.

The weighted minimum norm method that uses combined depth weighting and Laplacian smoothing, known as LORETA (low resolution brain electromagnetic tomography; Pascual-Marqui et al 1994), achieved the lowest localization error up to the present, among linear solutions. Yet, the method has non-zero error, but quite lower than the two previous methods.

## 7.2. eLORETA: optimal weights that produce exact localization

The regularized problem in Eq. 35 was presented from a "functional analysis" point of view. Alternatively, a Bayesian point of view renders the same formulation, where the quadratic functional in Eq. 35 is part of the posterior density, with:

**Eq. 38:** $\quad \Sigma_\Phi^{noise} = \alpha \mathbf{H}$

being the covariance matrix for the noise in the measurements, and:

**Eq. 39:** $\quad \Sigma_J = \mathbf{W}^{-1}$

being the "*a priori*" covariance matrix for the current density **J**.

Based on the linear relation in Eq. 14, extending it to include possible additive noise in the measurements, making use of Eq. 38 and Eq. 39, and assuming independence of neuronal activity and measurement noise, the covariance matrix for the electric potential is:

**Eq. 40:** $\quad \Sigma_\Phi = \mathbf{K}\Sigma_J \mathbf{K}^T + \Sigma_\Phi^{noise} = \mathbf{K}\mathbf{W}^{-1}\mathbf{K}^T + \alpha\mathbf{H}$

Based on the linear relation in Eq. 36, and making use of Eq. 40, the covariance matrix for the estimated current density is:

**Eq. 41:**
$$\begin{cases} \Sigma_{\hat{j}} = \mathbf{T}\Sigma_\Phi \mathbf{T}^T \\ \quad = \mathbf{T}\left(\mathbf{K}\mathbf{W}^{-1}\mathbf{K}^T + \alpha\mathbf{H}\right)\mathbf{T}^T \\ \quad = \mathbf{W}^{-1}\mathbf{K}^T \left(\mathbf{K}\mathbf{W}^{-1}\mathbf{K}^T + \alpha\mathbf{H}\right)^+ \left(\mathbf{K}\mathbf{W}^{-1}\mathbf{K}^T + \alpha\mathbf{H}\right)\left(\mathbf{K}\mathbf{W}^{-1}\mathbf{K}^T + \alpha\mathbf{H}\right)^+ \mathbf{K}\mathbf{W}^{-1} \\ \quad = \mathbf{W}^{-1}\mathbf{K}^T \left(\mathbf{K}\mathbf{W}^{-1}\mathbf{K}^T + \alpha\mathbf{H}\right)^+ \mathbf{K}\mathbf{W}^{-1} \end{cases}$$

When **W** is restricted to be a block-diagonal matrix, with the *j*-th block denoted as $\mathbf{W}_j \in \mathbb{R}^{3\times 3}$, for $j=1...N_V$, then the solution to the problem:

**Eq. 42:** $\quad \min_{\mathbf{W}} \left\| \mathbf{I} - \Sigma_{\hat{j}} \right\|^2 = \min_{\mathbf{W}} \left\| \mathbf{I} - \mathbf{W}^{-1}\mathbf{K}^T \left(\mathbf{K}\mathbf{W}^{-1}\mathbf{K}^T + \alpha\mathbf{H}\right)^+ \mathbf{K}\mathbf{W}^{-1} \right\|^2$

produces an inverse solution (Eq. 36 and Eq. 37) with zero localization error.

Zero localization error is defined in this study as follows: For a given point-test source anywhere in the solution space, with arbitrary orientation, compute the extracranial EEG/MEG measurements, give them to the linear inverse solution, threshold the inverse solution to the absolute maximum of the amplitude of the current density vector field, and compute as localization error the distance between the actual point-test source and the position of the absolute maximum.

This property has not been achieved by any previously published discrete 3D distributed linear solution.

Note that the covariance matrix for the estimated current density (Eq. 41) is not the resolution matrix of Backus and Gilbert.

The solution to the problem in Eq. 42 satisfies the following set of matrix equations:

Eq. 43: $\quad \mathbf{W}_j^2 = \mathbf{K}_j^T \left( \mathbf{K}\mathbf{W}^{-1}\mathbf{K}^T + \alpha\mathbf{H} \right)^+ \mathbf{K}_j$, for $j = 1...N_V$

where the matrix $\mathbf{K}_j$ is defined in Eq. 7.

The following simple iterative algorithm (in pseudo-code) converges to the block-diagonal weights $\mathbf{W}$ that solve the problem in Eq. 42 and equivalently satisfies Eq. 43:

1. Given the average reference lead field $\mathbf{K}$ and a regularization parameter $\alpha \geq 0$, initialize the block-diagonal weight matrix $\mathbf{W}$ as the identity matrix.
2. Set:

Eq. 44: $\quad \mathbf{M} = \left( \mathbf{K}\mathbf{W}^{-1}\mathbf{K}^T + \alpha\mathbf{H} \right)^+$

3. For $j = 1...N_V$ do:

Eq. 45: $\quad \mathbf{W}_j = \left[ \mathbf{K}_j^T \mathbf{M} \mathbf{K}_j \right]^{SymmSqrt}$

Comment: $\left[ \mathbf{K}_j^T \mathbf{M} \mathbf{K}_j \right]^{SymmSqrt}$ denotes the symmetric square root of the matrix $\left[ \mathbf{K}_j^T \mathbf{M} \mathbf{K}_j \right]$.

4. Go to step 2 until convergence (negligible changes in $\mathbf{W}$).

Finally, the block-diagonal matrix $\mathbf{W}$ produced by this algorithm should be plugged into the pseudoinverse matrix $\mathbf{T}$ (in Eq. 37). This is denoted as the *eLORETA* inverse solution.

## 7.3. eLORETA for EEG with known current density vector orientation, unknown amplitude

The average reference forward EEG equation (Eq. 14) is now written as:

Eq. 46: $\quad \mathbf{\Phi} = \mathbf{K}\mathbf{J} = \mathbf{K}\mathbf{N}\mathbf{L}$

with:

Eq. 47: $\quad \mathbf{J} = \mathbf{N}\mathbf{L}$

where $\mathbf{L} \in \mathbb{R}^{N_V \times 1}$ contains the current density amplitudes at each voxel, and $\mathbf{N} \in \mathbb{R}^{(3N_V) \times N_V}$ contains the outward normal vectors to the cortical surface at each voxel. Note that the columns of $\mathbf{N}$, denoted as $\mathbf{N}_j \in \mathbb{R}^{(3N_V) \times 1}$ for $j = 1...N_V$ are:

Eq. 48: $\quad \mathbf{N}_1 = \begin{pmatrix} \mathbf{n}_1 \\ \mathbf{o} \\ \mathbf{o} \\ \mathbf{o} \\ ... \\ \mathbf{o} \end{pmatrix} ; \mathbf{N}_2 = \begin{pmatrix} \mathbf{o} \\ \mathbf{n}_2 \\ \mathbf{o} \\ \mathbf{o} \\ ... \\ \mathbf{o} \end{pmatrix} ; \mathbf{N}_3 = \begin{pmatrix} \mathbf{o} \\ \mathbf{o} \\ \mathbf{n}_3 \\ \mathbf{o} \\ ... \\ \mathbf{o} \end{pmatrix} ; ... ; \mathbf{N}_{N_V} = \begin{pmatrix} \mathbf{o} \\ \mathbf{o} \\ \mathbf{o} \\ \mathbf{o} \\ ... \\ \mathbf{n}_{N_V} \end{pmatrix}$

where $\mathbf{o} \in \mathbb{R}^{3 \times 1}$ is a vector of zeros, and $\mathbf{n}_j \in \mathbb{R}^{3 \times 1}$ is the normal vector at the *j*-th voxel, i.e.:

Eq. 49: $\quad \mathbf{n}_j^T \mathbf{n}_j = 1$

In this section, $\mathbf{N}$ is assumed to be known.

The regularized, weighted minimum norm problem is:

Eq. 50: $$\min_{\mathbf{L}} \left[ \|\mathbf{\Phi} - \mathbf{KNL}\|^2 + \alpha \mathbf{L}^T \mathbf{W} \mathbf{L} \right]$$

where $\mathbf{W} \in \mathbb{R}^{N_V \times N_V}$ in this case denotes a given symmetric weight matrix, and $\alpha > 0$ denotes the regularization parameter.

The solution is linear:

Eq. 51: $$\hat{\mathbf{L}} = \mathbf{T}\mathbf{\Phi}$$

with:

Eq. 52: $$\mathbf{T} = \mathbf{W}^{-1}(\mathbf{KN})^T \left((\mathbf{KN})\mathbf{W}^{-1}(\mathbf{KN})^T + \alpha \mathbf{H}\right)^+$$

Following similar lines of reasoning as in the previous section, the covariance matrix for the electric potential is:

Eq. 53: $$\mathbf{\Sigma}_{\mathbf{\Phi}} = (\mathbf{KN})\mathbf{\Sigma}_{\mathbf{L}}(\mathbf{KN})^T + \mathbf{\Sigma}_{\mathbf{\Phi}}^{noise} = (\mathbf{KN})\mathbf{W}^{-1}(\mathbf{KN})^T + \alpha \mathbf{H}$$

where:

Eq. 54: $$\mathbf{\Sigma}_{\mathbf{L}} = \mathbf{W}^{-1}$$

is the "*a priori*" covariance matrix for the current density amplitudes **L**. In addition, the covariance matrix for the estimated current density is:

Eq. 55: $$\mathbf{\Sigma}_{\hat{\mathbf{L}}} = \mathbf{W}^{-1}(\mathbf{KN})^T \left((\mathbf{KN})\mathbf{W}^{-1}(\mathbf{KN})^T + \alpha \mathbf{H}\right)^+ (\mathbf{KN})\mathbf{W}^{-1}$$

When **W** is restricted to be a diagonal matrix, with the *j*-th element denoted as $W_j$, for $j = 1...N_V$, then the solution to the problem:

Eq. 56: $$\min_{\mathbf{W}} \|\mathbf{I} - \mathbf{\Sigma}_{\hat{\mathbf{L}}}\|^2 = \min_{\mathbf{W}} \left\| \mathbf{I} - \mathbf{W}^{-1}(\mathbf{KN})^T \left((\mathbf{KN})\mathbf{W}^{-1}(\mathbf{KN})^T + \alpha \mathbf{H}\right)^+ (\mathbf{KN})\mathbf{W}^{-1} \right\|^2$$

produces an inverse solution (Eq. 51 and Eq. 52) with zero localization error.

The solution to the problem in Eq. 56 satisfies the following set of equations:

Eq. 57: $$W_j = \sqrt{(\mathbf{KN})_j^T \left(\mathbf{KW}^{-1}\mathbf{K}^T + \alpha \mathbf{H}\right)^+ (\mathbf{KN})_j}, \text{ for } j = 1...N_V$$

where the vector $(\mathbf{KN})_j \in \mathbb{R}^{N_E \times 1}$ corresponds to the *j*-th column of $(\mathbf{KN})$.

The following simple iterative algorithm (in pseudo-code) converges to the diagonal weights **W** that solve the problem in Eq. 56 and equivalently satisfies Eq. 57:

1. Given the average reference lead field **K**, the cortical normal vectors **N**, and a regularization parameter $\alpha \geq 0$, initialize the diagonal weight matrix **W** as the identity matrix.
2. Set:

Eq. 58: $$\mathbf{M} = \left((\mathbf{KN})\mathbf{W}^{-1}(\mathbf{KN})^T + \alpha \mathbf{H}\right)^+$$

3. For $j = 1...N_V$ do:

Eq. 59: $$W_j = \sqrt{(\mathbf{KN})_j^T \mathbf{M}(\mathbf{KN})_j}$$

4. Go to step 2 until convergence (negligible changes in **W**).

Finally, the diagonal matrix **W** produced by this algorithm should be plugged into the pseudoinverse matrix **T** (in Eq. 52). This is denoted as the *e*LORETA inverse solution.

## 7.4. eLORETA for MEG with fully unknown current density vector field

This case follows the same derivations as given above for the case "EEG with fully unknown current density vector field".

The forward MEG equation has similar form to Eq. 14. For the MEG case, **Φ** would represent the magnetometer or gradiometer measurements, **K** would represent the magnetic lead field, and **J** is exactly the same current density vector field (common to both EEG and MEG).

In the MEG case, there is no reference electrode constant to be accounted for. The consequence is that the EEG regularization term $(\alpha \mathbf{H})$ appearing in some of the equations above (Eq. 37 to Eq. 44) must be changed to the MEG regularization term $(\alpha \mathbf{I})$, where **I** is the identity matrix.

In the case of spherical head models, care must be taken in the MEG case because only the tangential part of the current density vector field is non-silent. The same occurs in realistic head models, in areas that are quasi-spherical. This implies that all calculations at the voxel level have only rank=2 for MEG. Therefore, inverse and symmetric square-root matrix computations should be made via the singular value decomposition (SVD), ignoring the smallest eigenvalue if it is numerically negligible relative to the largest eigenvalue.

In particular, consider the algorithm involving Eq. 44 and Eq. 45. Note that Eq. 44 makes use of the inverse of the weight matrix, which consists of the inverses of all $3 \times 3$ block-diagonal submatrices. In the quasi-spherical MEG case, these submatrices have rank=2. Referring to Eq. 45, consider the SVD of the matrix of interest:

**Eq. 60:** $$\left[ \mathbf{K}_j^T \mathbf{M} \mathbf{K}_j \right] = \sum_{i=1}^{3} \lambda_i \mathbf{\Gamma}_i \mathbf{\Gamma}_i^T$$

where $\mathbf{\Gamma}_i \in \mathbb{R}^{3 \times 1}$ are the orthonormal eigenvectors, and $\lambda_1 \geq \lambda_2 \geq \lambda_3$ are the eigenvalues. Then Eq. 45 should be replaced by:

**Eq. 61:** $$\mathbf{W}_j = \left[ \mathbf{K}_j^T \mathbf{M} \mathbf{K}_j \right]^{SymmSqrt} = \begin{cases} \sum_{i=1}^{2} \sqrt{\lambda_i} \mathbf{\Gamma}_i \mathbf{\Gamma}_i^T \text{ , if } (\lambda_3/\lambda_1) < \varepsilon \\ \sum_{i=1}^{3} \sqrt{\lambda_i} \mathbf{\Gamma}_i \mathbf{\Gamma}_i^T \text{ , otherwise} \end{cases}$$

where $\varepsilon$ depends on the numerical precision of the calculations (typically $\varepsilon \leq 10^{-5}$).

Moreover, the inverse of $\mathbf{W}_j$ (see Eq. 61), which is needed in Eq. 44, and later on after convergence in Eq. 37 for the final inverse solution, should be calculated as the Moore-Penrose pseudoinverse (ignoring the smallest eigenvalue if it is numerically negligible relative to the largest eigenvalue).

Given these provisions and modifications, the discrete 3D distributed linear solution known as *e*LORETA is given by Eq. 36 and Eq. 37 with the weights defined by the solution the problem in Eq. 42, obtained with the algorithm specified by Eq. 44 and Eq. 45.

### 7.5. eLORETA for MEG with known current density vector orientation, unknown amplitude

This case follows the same derivations as given above for the case "EEG with known current density vector orientation, unknown amplitude".

The forward MEG equation in this case has similar form to Eq. 46. For the MEG case, $\mathbf{\Phi}$ would represent the magnetometer or gradiometer measurements, $\mathbf{K}$ would represent the magnetic lead field, $\mathbf{N}$ would contain the outward normal vectors to the cortical surface at each voxel (assumed known), and $\mathbf{L}$ contains exactly the same current density amplitudes (common to both EEG and MEG).

As explained previously, the EEG regularization term $(\alpha\mathbf{H})$ appearing in some of the equations above (Eq. 52 to Eq. 58) must be changed to the MEG regularization term $(\alpha\mathbf{I})$, where $\mathbf{I}$ is the identity matrix.

The existence of silent MEG sources might occur in practice, especially for quasi-radial sources in quasi-spherical head geometry. Care should be taken to exclude these possible silent sources from the solution space, even if this implies that there are missing cortical patches for the MEG solution space.

Given these provisions and modifications, the discrete 3D distributed linear solution known as *e*LORETA is given by Eq. 51 and Eq. 52 with the weights defined by the solution to the problem in Eq. 56, obtained with the algorithm specified by Eq. 58 and Eq. 59.

The general family of linear imaging methods is further extended to include data-dependent (adaptive) quasi-linear imaging methods, also with the exact, zero error localization property.

### 8. A family of discrete, 3D distributed quasi-linear imaging methods with exact, zero error localization: data-dependent (adaptive) methods

Formally, this family of methods is identical to the one defined by Eq. 15, except that now, instead of defining the parameter matrix $\mathbf{C}$ as a given, fixed matrix, it can, for example, be taken as the inverse covariance matrix for the measurements, i.e:

**Eq. 62:** $$\mathbf{C} = \left[ \frac{1}{N_K} \sum_{k=1}^{N_K} (\mathbf{\Phi}_k - \bar{\mathbf{\Phi}})(\mathbf{\Phi}_k - \bar{\mathbf{\Phi}})^T \right]^+$$

where the subscript "$k$" may index time or any other form of repeated measurements for the data. Note that in the case of MEG, $\mathbf{C}$ must be non-singular. In the case of EEG, $\mathbf{C}$ must be of rank $(N_E - 1)$, with its null eigenvector equal to a vector of ones (accounting for the reference constant). If the data so happens to be insufficient, i.e. $N_K \leq N_E$, or the data happens to be almost deterministic, resulting in a low rank matrix **C**, then the method will not have exact, zero error localization.

Eq. 62 corresponds to a single example illustrating the "adaptive" character of this family of methods. Any data dependent matrix **C** can be used, such as, for example, the squared inverse covariance matrix for the measurements.

Rigorously speaking, this method is not linear because the transformation depends on the data on which imaging is being carried out.

## 9. Conclusions

In Pascual-Marqui (1995, 1999, and 2002), the following arguments were used for selecting the best discrete, 3D distributed, linear tomography:
1. The aim of functional imaging is localization. Therefore, the best tomography is the one with minimum localization error.
2. In a linear tomography, the localization properties can be determined by using point-test sources, based on the principles of linearity and superposition.
3. If a linear tomography is incapable of zero error localization to point-test sources that are active one at a time, then the tomography will certainly be incapable of zero error localization to two or more simultaneously active sources.

Here we present a general family of linear imaging methods with exact, zero error localization to point-test sources.

We show that one particular member of this family, sLORETA (Pascual-Marqui 2002) has no localization bias in the presence of measurement noise and biological noise.

We introduce a new particular member of this family, denoted eLORETA. This is a genuine inverse solution and not merely a linear imaging method. We show that it has exact, zero error localization in the presence of measurement and structured biological noise. We derive and construct the method using two different approaches, and give practical algorithms for its estimation.

We present a general family of quasi-linear imaging methods that are data-dependent (adaptive). We also show that they are endowed with the exact, zero error localization property.

These results are expected to be of value to the EEG/MEG neuroimaging community.